\documentclass[12pt]{amsart}
\usepackage[initials]{amsrefs}
\usepackage{url}

\newcommand{\colonafter}[1]{#1:}

  \BibSpec{firstauthor}{
   +{}{ }{surname}
   +{,}{ }{initials}
   +{}{ }{jr}
  }
  \BibSpec{coauthor}{
   +{}{ and}{transition}
   +{}{ }{surname}
   +{,}{ }{initials}
   +{}{ }{jr}
  }
  \BibSpec{middleauthor}{
   +{,}{}{transition}
   +{}{ }{surname}
   +{,}{ }{initials}
   +{}{ }{jr}
  }
  \BibSpec{lastauthor}{
   +{,}{ and}{transition}
   +{}{ }{surname}
   +{,}{ }{initials}
   +{}{ }{jr}
    }

\BibSpec{miscellaneous}{%
+{}{\PrintAuthors}{author}
+{}{\PrintEditorsA}{editor}
+{.}{}{transition}
+{}{ \textit}{title}
+{,}{ }{type}
+{,}{ \EnglishInitialCaps}{booktitle}
+{,}{ Technical Report }{number}
+{,}{ }{series}
+{,}{ vol.~}{volume}
+{,}{ part~}{part}
+{.}{}{transition}
+{}{ \PrintYear}{date}
+{.}{}{transition}
+{}{ \colonafter}{place}
  +{}{ }{publisher}
  +{,}{ }{organization}
 +{,}{ }{institution}
+{,}{ \eprint} {eprint}
+{}{ \parenthesize}{status}
+{.}{}{transition}
+{}{ }{note}
+{.}{}{transition}
+{}{\SentenceSpace \ReviewList}{review}
+{}{ \parenthesize}{language}
+{}{ \url}{url}
}

   \BibSpec{article}{%
      +{}{\PrintAuthors}  {author}
      +{.}{}              {transition}
        +{}{ \PrintYear}           {date}
       +{.}{}              {transition}
    +{}{ }      {title}
       +{.}{}              {transition}
      +{}{ \textit}             {journal}
       +{.}{}              {transition}
      +{}{ \textbf}       {volume}
      +{} {\parenthesize} {number}
       +{:}{}              {transition}
      +{}{ }             {pages}
      +{.}{}              {transition}
      +{}{ }              {note}
  +{.}{}              {transition}
  +{} {\SentenceSpace \ReviewList}   {review}
  +{}{ \parenthesize}{language}
  +{}{}{part}    }

\BibSpec{book}{%
  +{}{\PrintAuthors}{author}
  +{}{\PrintEditorsA}{editor}
  +{.}{}{transition}
  +{}{ \PrintYear}{date}
  +{.}{}{transition}
   +{}{ \textit}{title}
  +{:}{ \textit}{subtitle}
  +{,}{ }{type}
  +{,}{ \EnglishInitialCaps}{booktitle}
  +{,}{ \PrintEdition}{edition}
  +{,}{ }{series}
  +{,}{ vol.~}{volume}
  +{,}{ part~}{part}
  +{.}{}{transition}
 +{}{ \colonafter}{place}
  +{}{ }{publisher}
  +{,}{ }{organization}
  +{}{ \parenthesize}{status}
  +{,}{ }{note}
  +{.}{}{transition}
  +{}{\SentenceSpace \ReviewList}{review}
  +{}{ \parenthesize}{language}
}

\title{A Comparison of Cryptography Courses}
\author{Joshua Holden}
\address{Department of Mathematics, Rose-Hulman Institute of Technology,
Terre Haute, IN 47803, USA}

\email{holden@rose-hulman.edu}


\begin{document}

\begin{abstract}  The author taught two courses on cryptography, one at
Duke University aimed at non-mathematics majors and one at Rose-Hulman
Institute of Technology aimed at mathematics and computer science majors.
Both tried to incorporate technical and societal aspects of cryptography,
with varying emphases.  This paper will discuss the strengths and
weaknesses of both courses and compare the differences in the author's
approach.
\end{abstract}

\maketitle

\section{Introduction}

This paper is a description, and to some degree a comparison, of two
courses on cryptography that I have taught.  One was a course in
``Cryptography and Society'' at Duke University, aimed at non-mathematics
majors and intended to explore both technical and societal aspects of
cryptography.  The other was a course in cryptography at my current school,
the Rose-Hulman Institute of Technology, an undergraduate engineering
college in Indiana.  This course was more technically oriented and aimed at
mathematics and computer science majors.

\section{``Cryptography and Society''}

During the fall of 2000, I taught a new course entitled ``Cryptography and
Society'' at Duke University.  This course was intended to introduce the
student to the basic ideas of modern cryptography and its applications.  It
was primarily aimed at non-mathematics majors; mathematics majors were
allowed to enroll but did not get credit towards their major.  The course
was suggested to me by my chair, Richard Hain, and was designed in part to
fit a new university requirement in Science, Technology, and Society, and
thus had a combination of technical and social topics.  I also made an
effort in the course to engage the students by bringing in examples from
their daily lives.  My approach to this course and many of the materials
used were strongly influenced by courses taught by Stephen Greenfield at
Rutgers University \cite{Greenfield}, Susan Landau at the University of
Massachusetts at Amherst \cite{Landau}, and William Pardon at Duke
University \cite{Pardon}.

The course was a seminar, and by university policy thus had an enrollment
cap of fifteen.  Fifteen students started the course, and thirteen
completed it.  All levels of undergraduate from first-year through senior
were present.  The students' majors included Biological Anthropology and
Anatomy, Biology, Economics, Mathematics, Political Science, Public Policy
Studies, and undeclared.  The published prerequisite was high school
algebra only.  Roughly half the students seemed to have no significant
amount of college mathematics, while the others had varying amounts up
through most of a mathematics major program.  Students were required to buy
four textbooks for the course:  Joseph Silverman's \emph{A Friendly
Introduction to Number Theory} \cite{Silverman}, Albrecht Beutelspacher's
\emph{Cryptology} \cite{Beutelspacher},  Whitfield Diffie and Susan
Landau's \emph{Privacy on the Line} \cite{DL},  and Simson Garfinkel and
Gene Spafford's \emph{Web Security \& Commerce} \cite{GS}. (Mathematics and
Computer Science professors who are used to assigning only one textbook for
the course should keep in mind the range of topics addressed and the fact
that humanities courses generally assign more textbooks than technical
ones.)  More information about the course may be found at my archived copy
of the course web page \cite{Math65S}.

Three main themes were addressed throughout the course: how modern
cryptographic protocols are implemented and their strengths and
weaknesses; how one encounters (and will increasingly encounter in the
future) cryptography in one's daily life; and the implications of
widespread use of cryptography in the digital age both for individuals
and for society.  In addition to studying the mathematics behind
modern cryptographic systems, we also examined the impact that the
invention of modern cryptographic systems has had and will have on
political, economic, philosophical, and sociological aspects of
society.  In order for the students to fully grasp this third theme,
of course, they needed to something about the mathematics of
cryptography, how it works and how it is used.

I tried in the course to schedule one day of mathematics, one
technical day, and one day of societal issues in each week rather than
splitting the course into blocks.  (The course was taught on Monday,
Wednesday, and Friday.)  Of course, this could not be followed
strictly for various reasons.  I also scheduled three guest speakers,
including someone from the Duke Medical Center, a Duke Law professor, and
the Mathematics Department system administrator.

The mathematical part of the course introduced basic number theory and work
with congruences, up through the Euler phi function and Euler's Theorem.
The goal was to get all of the mathematics necessary to understand the RSA
system of public-key encryption. There was also a short introduction to
finite fields as a prequel to the discussion of the Advanced Encryption
Standard (AES/Rijndael). My goals here were twofold. Firstly, I wanted to
make sure students understood the mathematics behind the cryptosystems.
Secondly I wanted them to see some of the basic ideas of number theory and
abstract algebra, notably the concept of numbers as things which could
behave differently depending on their context, that is, as an example of
abstract elements in a field or ring.

I did my best to introduce the mathematics gently, using \emph{A Friendly
Introduction to Number Theory}  supplemented by a few handouts. However,
the material quickly became quite hard for many of the students with less
mathematics background, and many seemed to feel that the textbook did not
help.  In retrospect, I think that the lack of mathematical prerequisites
may have given them the impression that the course was going to be less
mathematically intense than it turned out to be.  More mathematical
sophistication would have made the course easier, but I feel that it was
important to teach a course which students of all backgrounds could take.
Perhaps the answer was simply to be more up front (and more aware myself)
about how much mathematics there was going to be and how hard it was going
to get. Also, some topics, such as AES and elliptic curve cryptography
(ECC), required special mathematical material to be introduced just for
that topic.  Some students thought this wasn't well integrated into the
class.

Cryptosystems covered in the course included shift and affine ciphers, Hill
ciphers, linear feedback shift registers, DES, AES/Rijndael, RSA,
Diffie-Hellman key exchange, and ECC.  The more technical systems were
covered only in overview; notably DES, AES, and ECC. We also covered a
number of related protocols and technical matters which were not
cryptosystems, including digital signatures, message digests, subliminal
channels, zero-knowledge proofs, and systems for secure e-mail and secure
web browsing.  A certain amount of cryptanalysis was also discussed,
including the different types of information that a cryptanalyst might have
available.  There were three main goals for the selection of this group of
topics.  Firstly, I wanted to introduce a large selection of different
types of ciphers and other systems.  Secondly, I wanted ciphers which
reinforced the mathematics concepts I thought the students should know.
Thirdly, I wanted the students to see at least some ciphers which they
might encounter in their daily lives.

In addition, I tried to keep this aspect of the course grounded in what
cryptography could and could not be used for in practice, with an emphasis
on the limitations of algorithms in the face of the real world. The
textbook \emph{Cryptology} was used for this part of the course along with
a large number of handouts and web sites. Many of the students were quite
comfortable with programming and computers and this seemed to be the
easiest part of the course for them.

Discussions of the impact of cryptography on society were roughly
grouped into political, economic, and philosophical spheres.
Political implications that were discussed included the impact of
cryptography on law enforcement, on patent law, the possibility of
electronic voting, and the debate over personal privacy versus
national security.  Economic aspects centered around the startling
growth of electronic banking and electronic commerce in the last few
years, as well as the possibilities of legally binding digital
signatures and the use of ``digital cash''.  Philosophical and
sociological implications included outgrowths of many of the above
areas, especially the need to balance personal privacy with the public
interest.  In particular, we discussed law enforcement and the ability
of the government to eavesdrop on private communications, government
and corporate access to medical and banking records, and the ability
to track electronic purchases or prevent them from being tracked.

Readings from \emph{Privacy on the Line} were assigned to guide the
discussion of the political and philosophical issues, while that book and
\emph{Web Security \& Commerce} were used for economic issues.  A number of
handouts and web sites were also used in this part of the course. (A fairly
complete list of readings for the course may be found in the course
syllabus \cite{Math65Ssyllabus}.)  Classes for this part of the course were
generally conducted as open discussions, as opposed to the more
lecture-oriented classes in the rest of the course.  This was a new
experience for me as a mathematics professor, and as you might guess,
student interest waxed and waned depending on the topic, the point in the
course, the weather, and so on.  In general, however, I felt that the
discussions were useful and informative for everyone, myself included.
Students also seemed very positive about the discussions. Many (though not
all) students also liked the readings used in this part of the course.

Assignments for the course included weekly homework, a reading journal, a
term paper, an in-class midterm, and a take-home final. There was also a
small part of the grade based on classroom participation.  (I did not, in
the end, feel it necessary to penalize anyone for lack of participation.)
The first weekly homework was an ungraded student survey, which included
some basic information about the students, a pretest on some of the
mathematical content of the course, and an opinion survey with four
questions:
\begin{enumerate}
\item You should be able to copy any picture or music that you find on the
Internet.

\item Suitable government authorities should be able to have access to your
e-mail.

\item Personal financial and medical information should be totally secure
from every inquiry.

\item Math is needed to build rockets.  In order to make decisions about
business or government policy, most people don't need to know very much
math or even very much about what math can do.

\end{enumerate}
(These questions, and several others on the pretest, were taken from a
similar pretest used by Stephen Greenberg in his course.) I asked for a
rating of how strongly they agreed or disagreed and some justification.
There was a wide range of responses, which were also reflected in the way
the students responded during class discussions. In retrospect, it would
have been valuable to have re-done the survey at the end of class; I
suspect that student answers would have changed somewhat, though probably
not completely.  (It might also have been interesting to see how a
mathematics post-test went!)

Other weekly assignments included cipher problems to encode, decode,
or break, purely mathematical problems, and essays ranging from
paragraphs to 2--3 pages.  There was also an assignment in which they
had to use the computer program PGP to send, receive, and sign
encrypted e-mail messages.  Students seemed to enjoy the cipher
problems most, although some also got rather creative with their
essays.  I graded the essays on aspects of composition as well as
content; on the whole I was quite happy with the student writing.  The
mathematics problems seemed to cause the most problems.  I allowed
student to work in groups and had some difficulties early on with
students who were clearly copying.

The reading journal consisted of paragraph summaries collected every
week.  The articles were supposed to be related to cryptography
and published in a newspaper or newsmagazine during the previous month.
Most students in fact got these articles off of the web.  I did not
insist that the article originate in a print publication, but I did
try to make it clear that it should come from a professional news
organization rather than a weblog or similar site.  I also tried to
focus these journals on summarizing the article rather than
commentary, although I encouraged them to use their thoughts on the
articles in their essays (perhaps not as much as I could have).

The term paper assignment was a 1--2 page proposal followed by two drafts
of an 8--12 page paper.  The topic was left open; it could be oriented
towards mathematics, towards programming or software design, or towards
discussions of policy and social aspects of cryptography. I gave a list of
some examples.  Many of the students chose to write on historical topics
such as the Enigma cipher or the ADFGVX cipher. (The movie \emph{U-571},
which was tangentially related to the Enigma cipher, had recently premiered
at the time.)  The class focused largely on modern ciphers rather than
historical ones and some students, who were more historically minded, used
the term paper as an opportunity to bring more of their interests to the
course. This seemed quite appropriate to me, as I was interested in
historical topics myself but did not feel that there was sufficient time to
cover them in the course.

Both the in-class midterm exam and the take-home final were divided into
short answer questions, mathematics problems and ciphers, and an essay.
Short answer problems included:
\begin{enumerate}
\item Name two requirements mandated by HIPAA.

\item Name two public-key cryptosystems.  Name two symmetric-key
cryptosystems.  (Specify which are which!)

\item Define operations intelligence and give an example.

\item What is Kerckhoff's Principle?  Why is it important when defending
against known-plaintext attacks?

\item In a paragraph, describe three ways in which Rijndael is
    similar to DES and three ways in which Rijndael is different from
    DES. Be complete and thorough.

    \item In a paragraph, describe the difference between wiretaps,
    pen registers, and trap and trace.  What sort of authorization
    (under normal circumstances) does a law-enforcement officer need
    in order to use each of these techniques?

    \item In a paragraph, describe the differences between
    symmetric-key cryptosystems, message digests, and message
    authentication codes.  Include the name of a system that you
    might use for each of these, and the mode of use if appropriate.

    \item In a paragraph, describe three different uses we have made
    of ``cut-and-choose'' or ``sealed envelope'' techniques.  Why
    were the envelopes sealed in each case?  Why in each case did we
    choose some of the envelopes to open and others to remain sealed?

    \item In a paragraph, explain how digital watermarking is a form
    of steganography.  Then explain one form of steganography other
    than digital watermarking.
\end{enumerate}

The mathematics and cipher section of the final exam had two long problems:
\begin{enumerate}

\item
Your organization has just captured the secret plans (shown below) for the
nonlinear feedback shift register that the enemy has been using to encrypt
his communications.  Unfortunately, the nonlinear table of values was
smeared when your operative (Mallet) was forced to hide the plans in her
mouth.  However, Mallet was also able to learn that when the initialization
string 1001 was used, the the letters ``MA'' (in ASCII) were encrypted as
ciphertext 11010000 11011100.  Mount a known ciphertext attack by using
subtraction mod 2 to find the output of the nonlinear feedback shift
register.  Then work your way backwards, using subtraction mod 2 again
where necessary, to fill in the missing table.

\item
In this problem we will learn why we use modular arithmetic in the
Diffie-Hellman key exchange system.

\begin{enumerate}
    \item
You are Eve, and have captured Alice and Bob and imprisoned them.  You
overhear the following dialog.

Bob: Oh, let's not bother with the prime, it will make things easier.

Alice:  Okay, but we still need a base $s$ to raise things to.  How about
$s=3$?

Bob:  All right, then my result is 27.

Alice:  And mine is 243.

What is Bob's secret $b$ and Alice's secret $a$?  What is their secret
combined key?

\item Alice and Bob continue:

Bob: How should we use our new secret key?

Alice: Oh, that's easy; we'll use it as a one-time pad.  Divide the key
into 2-digit groups.  Use our cipher table (shown below) to change the
plaintext letters into numbers, and write the plaintext numbers left to
right under the key groups. Then add modulo 26 and convert it back to
letters.

Bob:  I can do that.  (Shouts) Eve is a TQYR!

    What did Bob say?

    \item Alice and Bob really should have reduced all their answers
    modulo some prime, such as $p=37$.  If they had done that, how
    would the exchange in part~(a) have gone?  What would have been
    their secret key?

    \item What advantages did you, as Eve,  have in part~(a) that Eve
    would not have in part~(c)?

    \item Would having a subliminal channel in part~(a) have helped Alice
    and Bob?  Why or why not?

    \end{enumerate}

\end{enumerate}
In retrospect, choosing two long problems seems to have been a mistake. For
each of the two problems, most students got either all of the points or
none of them, which did not allow for fine evaluation.  Also, there was not
much room for different ways of phrasing the answers, which made it easier
to grade but difficult to tell if there was copying going on.  (In the end
I am still not certain in the case of one incident whether copying
occurred.)

The midterm exam essay asked the student to write at least three good-sized
paragraphs on {one} of the following topics:
\begin{enumerate}
\item
The U.S. recently instituted new rules on export controls for cryptography.
What are the main points of the new rules?  Was the NSA for or against the
new rules?  Why?  What about the Department of Commerce?  Why?

or:

\item A major cryptographic patent recently was released by its holder.
What was it?  Why was it released?  In your opinion, should the patent have
been granted in the first place?  Why or why not?  What do you think the
impact will be of this patent passing into the public domain?  Why?
\end{enumerate}
The final exam essay asked the student to revisit one of the questions from
the opinion survey conducted during the first week of class.

\section{``Cryptography'' at Rose-Hulman}

More recently, I have been teaching a course entitled ``Cryptography''
at the Rose-Hulman Institute of Technology.  This was taught once
before I arrived at Rose-Hulman, by Prof.~David Mutchler of the
Computer Science department.  Prof.~Mutchler and I then team-taught
the course during the Spring Quarters of 2002 and 2003.  The course
was listed as a topics course in Computer Science during 2002, and was
cross-listed in both the Computer Science and Mathematics Departments
in 2003 (and will be in the future).  This course is primarily aimed
towards majors in the two departments, and could be can be counted
toward major programs in the departments in which it was listed.
The topics in this course were fairly similar to those in the Duke
course, but the tone was much more technically oriented.  We did try
to introduce some issues in the societal impact of cryptography.

In 2002, thirty-three students started the course and thirty-one completed
it.  (The official institute enrollment cap was thirty.)  In 2003,
twenty-five students started the course and twenty-one completed it.  (The
official institute enrollment cap was twenty-five.)  In both cases,
sophomores (including advanced standing first-years), juniors, and seniors
were all represented, with more of the more advanced students.  Majors were
predominantly computer science or computer science/\linebreak[1]mathematics
double majors, with a few each of computer engineering, mathematics, and
physics/\linebreak[1]computer science majors.  The published prerequisites
were one quarter of discrete mathematics (taken any time after the
completion of calculus) and two quarters of computer science courses.  The
discrete mathematics prerequisite was waived in one case, which proved not
to be a good idea. Students were required to buy two textbooks for the
course: Williams Stallings' \emph{Cryptography and Network Security}
\cite{Stallings} and Stephen Levy's \emph{Crypto: How the Code Rebels Beat
the Government --- Saving Privacy in the Digital Age} \cite{Levy}. (Once
again, two textbooks were necessary because of the incorporation of both
technical and non-technical subjects.) More information about the course
may be found at the course web page \cite{Math479}.

This course focused primarily on the mathematical background and practical
implementation of modern cryptographic protocols.  Although not formally a
seminar, we tried to encourage an attitude where learning interesting
material was primary and grades were secondary. To this end, there were no
tests in the course and assignments were designed to allow students to
focus on topics they found most interesting.  As in the course at Duke, the
mathematics and the technical details were intertwined with each other and
with some discussion of societal issues.  A typical four-class week might
have one day concentrating on mathematics, two on cryptosystems or other
protocols, and one on (recent) historical or societal issues.

The mathematical part of the course was very similar to the Duke
course.  The students were somewhat more mathematically sophisticated
and had an easier time absorbing the topics, although the topics
themselves were still mostly new to all but a few of the students.  We
spent more time on finite fields and elliptic curves than I did in the
Duke course and thus were able to go into more detail with AES and
elliptic curve cryptography.  The goals of the mathematical part of
the course were similar to those at Duke, although we also aimed to
impart a degree of practical knowledge that the students might need in
their future careers.

Cryptosystems covered in the course included shift and affine ciphers, Hill
ciphers, Simplified DES, DES, Simplified AES, AES/Rijndael, RSA,
Diffie-Hellman key exchange, and elliptic curve cryptography.  The more
technical background of the students allowed us to cover DES and AES in
quite a bit more detail than I could at Duke.  We used the simplified
versions of these two algorithms created by Ed Schaefer (Santa Clara
University) and his students \cites{SDES, SAES} in order to give our
students hands-on experience with these ciphers.  We also covered digital
signatures, subliminal channels, zero-knowledge proofs, and a discussion of
the information-theoretic idea of perfect secrecy.  The mathematical and
technical aspects of the course used  \emph{Cryptography and Network
Security} as a textbook, along with some handouts and web sites.  Some
students seemed to think that a cryptography reference book would be more
useful than a traditional textbook, and we may recommend one in the future,
although I do not think we will require students to buy it. (Bruce
Schneier's \emph{Applied Cryptography} \cite{Schneier} is one possibility.)

Discussions of cryptography and society revolved around readings from
\emph{Crypto: How the Code Rebels Beat the Government --- Saving Privacy in
the Digital Age}. This is a popular, if somewhat sensational, account of
the development of modern cryptography from roughly 1970 to the present.
The story told in the book includes the development of many of the key
ideas and protocols we discussed in the course.  It also explores the
reactions of government, business, and society to these developments,
giving us a handle on which to hang class discussions on some of the same
political, economic, and philosophical ideas covered in the Duke course.
The book is certainly more lively than the books used in the Duke course.
However, the overtly political tone put off some of the students in the
course.

Assignments for the course included weekly homework, an oral report, and a
research proposal.  Students were also given points for readings and class
discussions on \emph{Crypto}.  In the 2002 version of the course students
were not given enough weekly homework and I think this was detrimental to
their learning, especially of the more advanced topics.  In the 2003
version students were given homework slightly less than weekly (six times
during the ten-week course).  Each homework was divided into
``mathematics-inspired problems'' and ``computer science-inspired
problems''.  The ``mathematics-inspired problems'' ranged over
mathematical, algorithmic, and protocol-based topics but were all intended
to be solved with paper and pencil and perhaps the assistance of a small
computer program or a computer algebra system such as Maple.  Many, but not
all, were taken from \emph{Cryptography and Network Security}.  The
``computer science-inspired problems'' were slightly larger programming
projects and included implementations of various cryptosystems and
protocols discussed in the class, investigations of cryptanalytic
techniques, and the gathering of empirical evidence for various theorems
and conjectures related to the course.

More homework problems were made available than students could possibly be
expected to do.  Each was assigned a number of points, and a certain
minimum number of points was required on each part (``mathematics'' and
``computer science'') of each homework assignment for a passing grade.
Students could increase their grade beyond the minimum by earning points on
any part of any assignment, without restriction. Our object was to
encourage students to find problems which interested them, while still
making sure they had some basic understanding of all areas of the course.
Most students found the mathematics part of the homework more difficult
than the computer science part, which was not surprising since most were
computer science majors.

There were also several in-class projects which gave students a
hands-on experience with some of the more important cryptosystems.
The students worked through worksheets in small groups with assistance
from the instructors.  Most of these could also be turned in for
homework credit.  The topics included Simplified DES, the
cryptanalysis of Simplified DES, Simplified AES,
and RSA. (The RSA worksheet focused on the algorithms for fast
exponentiation and fast probabilistic primality testing.)

The oral presentation was based on a technical article on cryptography
chosen by the students with help from the instructors.  Suggested sources
included \emph{Cryptologia}, the \emph{Journal of Cryptology}, and the
proceedings of CRYPTO, EUROCRYPT, ASIACRYPT, and similar conferences.
Presentations were 20 minutes long and were intended to show that the
student understood the basic technical ideas of the article and could
communicate them in a way that impressed their importance upon the
audience.  Students were encouraged to use PowerPoint or similar
presentation software.  The difficulty of the material varied widely,
depending on the level of the student.  We took this into account to a
degree when grading the article, but more emphasis was placed on making
sure the student understood the material, regardless of difficulty. Quantum
cryptography was a popular subject during the 2002 version of the course.
Primality testing was very popular during the 2003 course, with three
students collaborating on a series of presentations about the AKS
deterministic primality test.  Image-based steganography was a popular
subject during both years.  Quality of presentations varied quite a bit, to
the degree that we had to ask some of the students in 2003 to repeat their
presentations in order to get credit.

The research proposal was a 2--5 page paper which was intended to give
students a chance to think about open problems and a taste of academic
research in the field of cryptography.  Students were asked to formulate
and clearly state an open problem, explain why solving the problem would be
valuable, and propose some promising directions for solving the problem.
The students were told that the proposal would be considered a success if
the student could convince the instructors that he or she had promising
ideas that had a reasonable chance of contributing to the solution of the
problem.  Students were encouraged to get their open problems from the
presentations of other students, although this was not strictly required.
Students were discouraged from choosing a famous unsolved problem in the
field, e.g., a proof of the security of RSA. Students were not required to
carry out any of the actual research, of course.

Some proposals made by students included designing switching networks for
use with quantum cryptography, the use of ``cwatsets'' \cite{cwatsets} in
symmetric-key cryptography, finding new ways to attack the one-time pad
statistically, detecting cheating in translucent cryptography systems, and
detecting image steganography using statistical methods. This was the first
time most students had ever been asked to consider finding a significant
problem of their own to investigate, and many students had to be encouraged
to apply some creative and original thought. Also, few had any experience
judging what made an idea likely to work in solving a research problem.  In
the end, some of the proposals were rather pedestrian, but some included
some genuinely creative ideas. On the whole we thought the experience was
very valuable for students, especially those who were nearing graduation
and the ``real world''.

The final portion of the grade was based on the readings from
\emph{Crypto} and the accompanying classroom discussions, which were
conducted in a similar fashion to the Duke course.  At the beginning
of each discussion, students were asked to do a ``two-minute essay''
answering two easy questions from the reading.  Often, the questions
asked the students what found most interesting on a certain topic.
These ``essays'' were use to judge whether the students had done the
reading and also sometimes to stimulate the classroom discussions.
Students were also given points for being present on the day of the
discussion.  Not all students participated actively in the
discussions, but many contributed comments and there was a certain
amount of heated debate.  In general, students in the Rose-Hulman course did
not seem as inclined to argue about these topics as the students in
the Duke course.  This was probably due to a combination of the
difference in the course and the difference in the student body as a
whole.

\section{Conclusion}

One of the striking things about these two courses is how a fairly similar
set of material can produce two entirely different courses depending on the
emphasis of the instructor and the abilities and interests of the students.
Obviously, the stronger technical background of the Rose-Hulman students
allowed a greater amount of depth in the technical areas of the course.  On
the other hand, the less technically oriented non-mathematics majors at
Duke were perhaps better able to come to grips with the social aspects and
implications of cryptography which we tried to bring out in both courses.
One of the things that I hoped to achieve in the Duke course was to
demonstrate that there was value in bringing technical material on
cryptography to students with a limited background but a strong interest.
This was not as successful as I would have liked, and a comparison of my
experiences in the two courses makes it clear that a stronger technical
background does lead to a better understanding of the material.  This is
perhaps no surprise.  I still feel, however, that introducing the Duke
students to technical aspects of cryptography also enhanced their
understanding of the non-technical aspects.

\section*{Acknowledgments}

I would like to thank my department chairs, Richard Hain at Duke and Allen
Broughton at Rose-Hulman, for giving me the opportunity to teach these
courses.  I also would like to thank Stephen Greenfield, Susan Landau,
William Pardon, and of course my co-teacher David Mutchler for showing me
how to teach a course in cryptography and providing materials which were
invaluable to me in teaching these courses.  I would like very much to
thank all of the students who have taken my courses for their
contributions, comments, and enthusiasm when appropriate.  Finally, I would
like to thank Brian Winkel for suggesting that I write this article, and
for his helpful comments on an early draft.

\section*{Biographical Sketch}

Joshua Holden is currently in his third year in the Mathematics Department
of Rose-Hulman Institute of Technology, an undergraduate engineering
college in Indiana.  He received his Ph.D. from Brown University in 1998
and held postdoctoral positions at the University of Massachusetts at
Amherst and Duke University.  His research interests are in computational
and algebraic number theory and in cryptography.  His teaching interests
include the use of technology in teaching and the teaching of mathematics
to computer science majors, as well as the use of historically informed
pedagogy.  His non-mathematical interests currently include science
fiction, textile arts, and choral singing.

\end{document}